\documentclass[aps,prb,showpacs,twocolumn,superscriptaddress]{revtex4} 

\usepackage[]{graphicx,color}
\usepackage[version=3]{mhchem}

\begin{document}

\title{
Monoclinic \ce{YFe_{12}} phases predicted from first principles
}



\author{Takahiro Ishikawa}%
 \email{ISHIKAWA.Takahiro@nims.go.jp}
 \affiliation{%
 ESICMM, National Institute for Materials Science, 1-2-1 Sengen, Tsukuba, Ibaraki 305-0047, Japan
 }%
\author{Taro Fukazawa}%
 \affiliation{%
 CD-FMat, National Institute of Advanced Industrial Science and Technology, 1-1-1 Umezono, Tsukuba, Ibaraki 305-8568, Japan
 }%
 \affiliation{%
 ESICMM, National Institute for Materials Science, 1-2-1 Sengen, Tsukuba, Ibaraki 305-0047, Japan
 }%
\author{Takashi Miyake}%
 \affiliation{%
 CD-FMat, National Institute of Advanced Industrial Science and Technology, 1-1-1 Umezono, Tsukuba, Ibaraki 305-8568, Japan
 }%
 \affiliation{%
 ESICMM, National Institute for Materials Science, 1-2-1 Sengen, Tsukuba, Ibaraki 305-0047, Japan
 }%

\date{\today}

\begin{abstract}
We searched for stable crystal structures of \ce{YFe_{12}} using a crystal structure prediction technique based on a genetic algorithm and first-principles calculations. We obtained two monoclinic $C2/m$ structures as metastable phases that are different from the well-known \ce{ThMn_{12}} structure.
These two phases have advantages in their magnetism over the \ce{ThMn_{12}} structure: 
The total magnetization $M$ is increased from 25.6\,$\mu_{\text{B}}$/f.u. up to 26.8\,$\mu_{\text{B}}$/f.u. by the transformations. We also calculated Curie temperature $T_{\text{C}}$ for these structures within the mean-field approximation and predicted the increase of $T_{\text{C}}$ from 792\,K up to 940\,K, which is mainly caused by the increase of intersite magnetic couplings within the distance of 2.3--3.1\AA. The similar enhancements of $M$ and $T_{\text{C}}$ are also obtained in the pseudo-binary system \ce{Y(Fe_{1-$x$}Co_{$x$})_{12}} with $x$ of 0--0.7. 
\end{abstract}


\maketitle


\section{Introduction}

Modern high-performance permanent magnets are rare-earth magnets, whose main phases mainly consist of 3d transition metals (Fe and/or Co) and rare-earth elements. High Fe/Co concentration gives rise to high magnetization ($M$) and high Curie temperature ($T_{\text{C}}$), and rare earths are a source of high magnetocrystalline anisotropy which is essential for high coercivity. For example, neodymium magnets are the strongest type of permanent magnet commercially available, and its main phase is formed by \ce{Nd_{2}Fe_{14}B} compound~\cite{Sagawa1984}. 
The $M$ value is expected to be further increased if a compound having more iron content than \ce{Nd_{2}Fe_{14}B} is used. 

For this reason, \ce{$R$Fe_{12}} ($R$ = rare earth element) compounds with the \ce{ThMn_{12}} structure have been regarded as
potential candidates for strong magnet compounds~\cite{Ohashi1988-JApplPhys,Buschow1988,Yang1988}. 
A few years ago, thin films of \ce{NdFe_{12}N_{$x$}} and \ce{Sm(Fe_{1-$x$}Co_{$x$})_{12}} have been fabricated by the epitaxial growth on W- and V-buffered \ce{MgO}(001) substrates, and it has been clarified that they have higher $M$, $T_{\text{C}}$, spontaneous magnetization, and anisotropy field than \ce{Nd_{2}Fe_{14}B}~\cite{Hirayama2015,Hirayama2017}. 
However, it has long been known that the 
\ce{$R$Fe_{12}} compounds are thermodynamically unstable in a bulk form and are stabilized by partial substitution of the third element $X$
for Fe,
\textit{i.e.} \ce{$R$(Fe_{1-$x$}$X$_{$x$})_{12}} ($X$ = Al, Si, Ti, V, Cr, Nb, Mo, and W) , whereas the $M$ value is decreased with the increase of $x$~\cite{Felner1983,Ohashi1987,Ohashi1988,DeMooij1988,Fuquana2005,Miyake2014,Harashima2016}.  
On the basis of this background, a search for the best stabilizing element has been a major approach in 
development of high-performance magnets. 

We are searching for novel Fe-rich magnetic compounds using first-principles calculations. 
In the present study, we apply another approach, the exploration of novel crystal structures for \ce{$R$Fe_{12}} 
showing higher performance as magnetic compounds than the \ce{ThMn_{12}} structure.
To achieve this, we use a scheme of genetic algorithms (GA),
which is a heuristic approach 
to solve problems using 
mechanisms inspired by the biological evolution, ({\it e.g.}, mating, mutation,
selection, inheritance, {\it etc.}).
The GA schemes have been applied to search for
stable and metastable crystal structures and have
succeeded in several types of materials.~\cite{Deaven1995,Bush1995,Woodley1999,Woodley2004,Oganov06,Abraham2006,Abraham2008,Ishikawa2019,Ishikawa2020-CH}. 
We here focus on \ce{YFe_{12}} because
Y has no $f$ electron in its ground electronic configuration, which is favorable for theoretical treatment, and \ce{YFe_{12}} with the \ce{ThMn_{12}} structure is experimentally obtained in multi-phases by rapid quenching method~\cite{Suzuki2017}. 
First we present stable structures of \ce{YFe_{12}} obtained by our GA scheme combined with first-principles calculations. 
Then, we show the values of $M$ and $T_{\text{C}}$ 
calculated within the mean field approximation
to compare the performance as magnet compounds with those of \ce{ThMn_{12}}. 

\section{Computational details}

Details of our GA structure search are shown in Ref. \onlinecite{Ishikawa2019}. 
In this study, first (i) we prepared a population consisting of 20 crystal structures, which are generated randomly. Those structures are optimized using a first-principles package, and ranked according to the total energy $E$. 
Next, (ii) new structures are created by applying evolutionary operators and performing the structural optimization: eight structures by ``mating'' (making a slab structure from two structures randomly selected) and 12 structures by ``mutation''  
(distorting the lattice or permuting the atomic positions of a structure randomly selected). 
Then, (iii) the population for the next generation is constructed by inheriting four elite structures with the lowest $E$ values at the previous generation, ranking all the 24 structures by $E$, and eliminating four unstable structures with the highest $E$ values. By repeatedly performing (ii) and (iii), 
energetically stable structures are obtained.

We combined our structure search code with the Quantum ESPRESSO (QE) code~\cite{QE} to perform the structural optimizations. 
We used calculation cells including 1-4 formula units of \ce{YFe_{12}}. 
The generalized gradient approximation by 
Perdew, Burke and Ernzerhof~\cite{PBE} was used for the exchange-correlation functional, and 
the Rabe-Rappe-Kaxiras-Joannopoulos ultrasoft pseudopotential~\cite{RRKJ90} was employed. 
The $k$-space integration over the Brillouin zone (BZ) was carried out 
on a 4 $\times$ 4 $\times$ 4 grid, and the energy cutoff was set at 80\,Ry for the wave function and 640\,Ry for the charge density. 
After obtained the stable structures for each of the different number of the formula units by GA, 
we compared the energy differences among them by increasing 
the number of $k$ points to 8 $\times$ 8 $\times$ 8. 

For the obtained structures, we calculate the intersite magnetic couplings using Liechtenstein's method~\cite{Liechtenstein87}. 
For this purpose, we used AkaiKKR\cite{AkaiKKR},
a first-principles program of Korringa-Kohn-Rostoker (KKR) Green's
function method, within the local density approximation.
The $T_{\text{C}}$ value is evaluated from a classical spin model within the mean-field approximation. 
Other computational details are same to the settings
in Ref.~\onlinecite{Fukazawa18}.

\section{Results}

In this study, we searched for not only the most stable structure but also metastable structures with instability energy $\Delta E$ less than 50\,meV/atom. 
This tolerance is associated with the approximations and the omission of temperature effects in first-principles calculations~\cite{Wu2013,Hinuma2016}, and the possibility of the stabilization by the inclusion of the third elements, such as the cases of \ce{Nd_{2}Fe_{14}B} and \ce{$R$(Fe_{1-$x$}$X$_{$x$})_{12}}. 
Applying the GA structure search to \ce{YFe_{12}}, we obtained 
the well-known \ce{ThMn_{12}} structure with tetragonal $I4/mmm$ as the most stable one and two novel monoclinic $C2/m$ structures as the second and third most stable ones. 
Hereafter we call $C2/m$ with $\Delta E = 38.9$\,meV/atom (second most stable structure) ``type-I'' and that with $\Delta E = 43.0$\,meV/atom (third most stable one) ``type-II''. 
These $\Delta E$ values correspond to temperatures of 450--500\,K. 
The structure parameters are listed in Table \ref{structureparameter}. 
\begin{figure*}
\includegraphics[width=18cm]{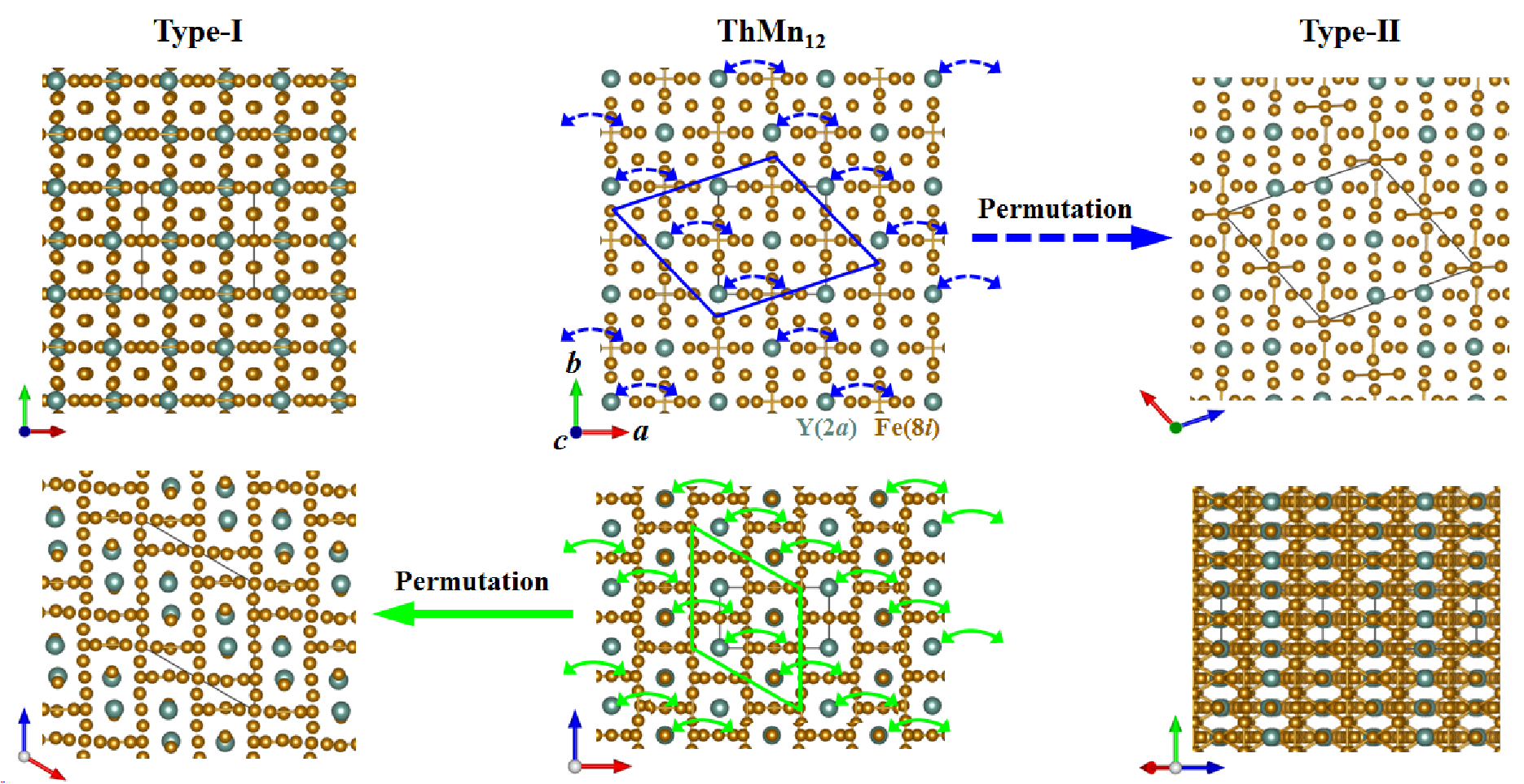}
\caption{\label{Fig-structure} 
(Color online) Monoclinic $C2/m$ structures, type-I and type-II, obtained via the partial permutations of atomic positions between Y at $2a$ site and Fe at $8i$ site of the \ce{ThMn_{12}}-type structure. 
Large and small balls represent Y and Fe atoms, respectively. 
The permuted atoms are shown by solid arrows for type-I and broken arrows for type-II. 
The parallelograms in \ce{ThMn_{12}} show the unit cells of type-I and type-II.
Crystal structures were drawn with VESTA~\cite{VESTA}.}
\end{figure*}
\begin{table}
\caption{\label{structureparameter}
Space group (SG), cell parameters, and atomic positions of metastable type-I and type-II structures with respect to \ce{YFe_{12}}.}
\begin{ruledtabular}
\begin{tabular}{ccll}
& SG & Cell (\AA, $^{\circ}$) & Atomic position \\ \hline
Type-I &$C2/m$ & $a$ 9.8712 & Fe $4i$ 0.8634 0 -0.0609 \\ 
& & $b$ 8.1400 & Fe $2b$ 0 0.5 0\\
& & $c$ 9.6306 & Fe $8j$ 0.0136 0.2475 0.3792\\
& & $\beta$ 119.78 & Fe $8j$ 0.4875 0.2490 0.8707\\
& & & Fe $4i$ 0.0071 0 0.2499\\
& & & Fe $8j$ 0.7398 0.1491 0.6639\\
& & & Fe $8j$ 0.2448 0.1682 0.8918\\
& & & Fe $4i$ 0.5955 0 0.8030\\
& & & Fe $2d$ 0 0.5 0.5\\
& & & Y $4i$ 0.7614 0 0.3839\\ \hline
Type-II &$C2/m$ & $a$ 11.8702 & Fe $4i$ 0.8777 0 0.7402 \\
& & $b$ 4.7233 & Fe $8j$ 0.1922 0.2523 0.6241 \\
& & $c$ 13.0295 & Fe $4i$ 0.6627 0 0.1934\\
& & $\beta$ 113.76 & Fe $4i$ 0.0990 0 0.7312\\ 
& & & Fe $4i$ 0.1887 0 0.4631\\
& & & Fe $4i$ 0.2647 0 -0.0389\\
& & & Fe $2b$ 0 0.5 0\\
& & & Fe $2c$ 0 0 0.5\\
& & & Fe $4i$ 0.1047 0 0.0627\\
& & & Fe $8j$ 0.4453 0.2483 0.1287\\
& & & Fe $4i$ 0.3041 0 0.1989\\
& & & Y $4i$ 0.4491 0 0.6247\\
\end{tabular}
\end{ruledtabular}
\end{table}

Figure \ref{Fig-structure} shows the comparison among the \ce{ThMn_{12}}, type-I, and type-II structures. Both the two $C2/m$ structures are achieved via partial permutation 
between Y at the $2a$ site and Fe at the $8i$ site of the \ce{ThMn_{12}} structure along the direction parallel to the $a$ axis. 
The permutation of Y and Fe indicated in the figure is repeated along the $b$ ($c$) axis of \ce{ThMn_{12}} with the interval of the cell-edge length.
The nearest Y-Y distance is decreased from 6.41 to 4.90 and 3.90\AA\, by the transformation from \ce{ThMn_{12}} into type-I and type-II, respectively.

\begin{figure}
\includegraphics[width=8.4cm]{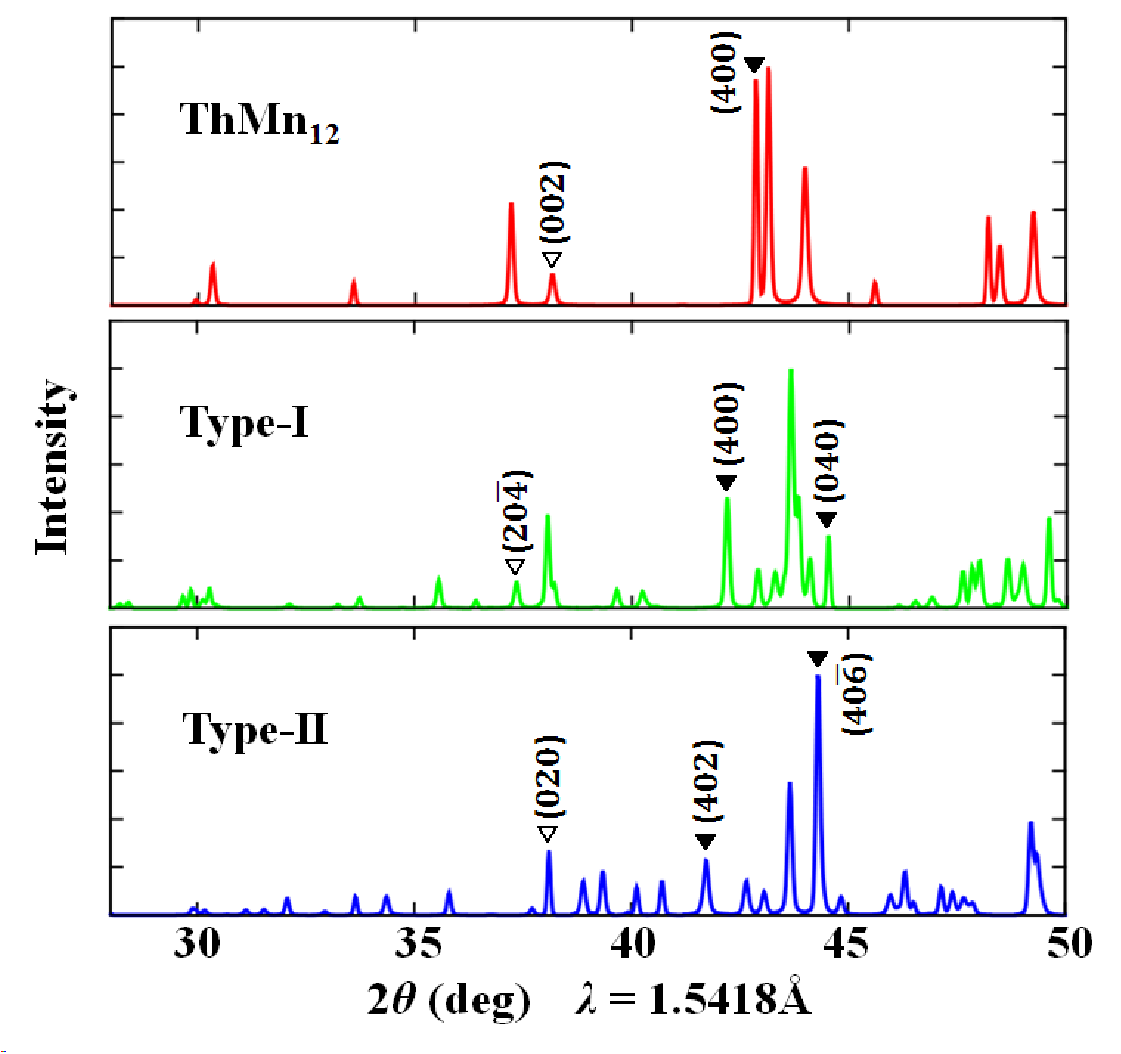}
\caption{\label{Fig-XRD} 
(Color online) X-ray diffraction patterns of the \ce{ThMn_{12}} structure with tetragonal $I4/mmm$, and the type-I and type-II structures with monoclinic $C2/m$, simulated by RIETAN-2000~\cite{Izu00}. 
The diffraction peaks concerned with the (002) and (400) peaks of \ce{ThMn_{12}} are indicated by open and closed inverted-triangles, respectively.
}
\end{figure}
Figure \ref{Fig-XRD} shows x-ray diffraction (XRD) patterns, simulated by RIETAN-2000~\cite{Izu00}, with respect to the \ce{ThMn_{12}}, type-I, and type-II structures. 
The wavelength $\lambda$ was set at 1.5418\AA. 
Here we focus on how the (002) and (400) diffraction peaks of \ce{ThMn_{12}}, which are clearly observed by the experiments~\cite{Hirayama2017,Hadjipanayis2020}, are varied by the transformation into type-I and type-II. 
In the XRD patterns of type-I and type-II, the peaks concerned with the (002) and (400) peaks are indicated by open and closed inverted-triangles, respectively. 
For type-I, the position of the (20$\bar{4}$) peak, which corresponds to the (002) peak at $2\theta = 38.49^{\circ}$ in \ce{ThMn_{12}}, is shifted to 37.35$^{\circ}$, and the (400) peak at 42.82$^{\circ}$ in \ce{ThMn_{12}} splits into the (400) peak at 42.19$^{\circ}$ and the (040) peak at 44.52$^{\circ}$ because the $a$ and $b$ axes get to be inequivalent due to the transformation from the tetragonal $I4/mmm$ to the monoclinic $C2/m$. 
For type-II, the (020) peak emerges at 38.10$^{\circ}$, which is almost unchanged from that of the (002) peak in \ce{ThMn_{12}}, and the (400) peak in \ce{ThMn_{12}} splits into the (402) peak at 41.73$^{\circ}$ and the (40$\bar{6}$) peak at 44.32$^{\circ}$. The (40$\bar{6}$) peak has the largest intensity of all the diffraction peaks in type-II.  
In addition, many small peaks appear in the XRD patterns of type-I and type-II due to the lowering of crystalline symmetry. 
See Fig. S1 in Supplemental Material for the $(hkl)$ values of the peaks~\cite{SM_XRD}. 

\begin{figure}
\includegraphics[width=8.6cm]{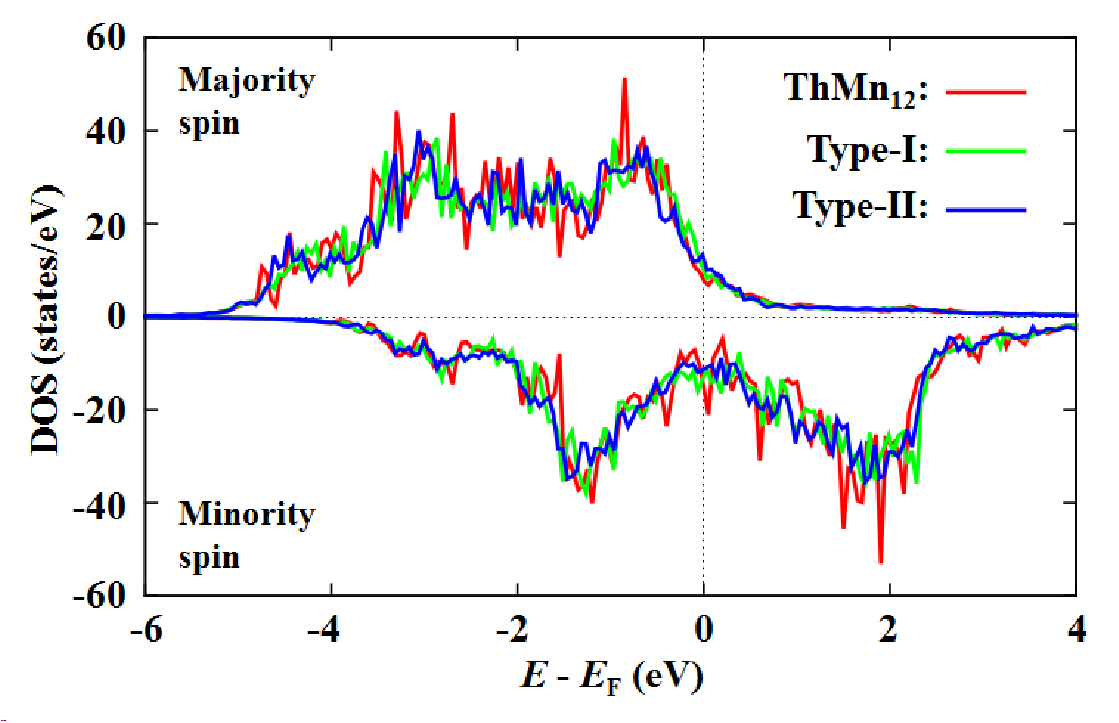}
\caption{\label{Fig-DOS} 
(Color online) Density of states (DOS) for the 3$d$ states of Fe atoms as a function of energy relative to the Fermi level ($E_{\text{F}}$).}
\end{figure}
We calculated the density of states (DOS) for the \ce{ThMn_{12}}, type-I, and type-II structures
to examine their electronic structures.
Figure \ref{Fig-DOS} shows the partial density of the 3$d$ states at the Fe sites for the three structures. 
The type-I and type-II structures have two formula units in the primitive cell, whereas 
\ce{ThMn_{12}} has a formula unit. 
To make the comparison easier, DOS of \ce{ThMn_{12}} was doubled in this figure. 
\ce{ThMn_{12}} has a large minority-spin DOS at the Fermi level ($E_{\text{F}}$) due to the appearance of a van Hove singularity~\cite{vanHove1953} slightly above $E_{\text{F}}$. 
The DOS is broadly smoothened with the crystal symmetry lowered by the transformation from tetragonal \ce{ThMn_{12}} into monoclinic type-I and type-II. The singularity then disappears, and the minority-spin DOS at $E_{\text{F}}$ decreases. 

\begin{table*}
\caption{\label{data_comparison}
Instability energy $\Delta E$, volume $V$, total magnetization $M$, and Curie temperature $T_{\text{C}}$ for the \ce{ThMn_{12}}, type-I, and type-II structures in \ce{YFe_{12}} and \ce{YCo_{12}}.}
\begin{ruledtabular}
\begin{tabular}{ccccccc}
&  & $\Delta E$ & $V$ & \multicolumn{2}{c}{$M$} & $T_{\text{C}}$ \\
&  & (meV/atom) & (\AA$^{3}$/f.u.) & ($\mu_{\text{B}}$/f.u.) & (T) & (K) \\ \hline
\ce{YFe_{12}}&\ce{ThMn_{12}} & 0 & 166.9 & 25.6 & 1.79 & 792 \\
&Type-I & 38.9 & 167.9 & 26.8 & 1.86 & 830 \\
&Type-II & 43.0 & 167.2 & 26.1 & 1.82 & 940 \\ \hline
\ce{YCo_{12}}&\ce{ThMn_{12}} & 0 & 158.1 & 19.2 & 1.42 & 1280 \\
&Type-I & 27.1 & 159.0 & 18.8 & 1.38 & 1290 \\
&Type-II & 37.3 & 159.5 & 19.0 & 1.39 & 1282 \\
\end{tabular}
\end{ruledtabular}
\end{table*}
We calculated the total magnetization $M$ and Curie temperature $T_{\text{C}}$ of the type-I and type-II structures and compared them with those of \ce{ThMn_{12}}. 
The results are listed in Table \ref{data_comparison}. the $M$ value is increased from 25.6 to 26.8 and 26.1\,$\mu_{\text{B}}$/f.u. by the transformation from \ce{ThMn_{12}} into type-I and type-II, respectively. 
The $T_{\text{C}}$ value is also increased from 792 to 830\,K by the transformation into type-I, and is more largely increased to 940\,K by the transformation into type-II. 
Although the mean-field approximation tends to overestimate $T_{\text{C}}$, the differences of theoretical $T_{\text{C}}$ values among magnet compounds have been found to be qualitatively consistent with those in experiments~\cite{Fukazawa17,Fukazawa18,Fukazawa2019-Tc}. Therefore, we consider the degree of the enhancement in $T_{\text C}$ is realistic.

We performed a similar calculation for \ce{YCo_{12}}, substituting Co for Fe. 
The monoclinic structures of \ce{YCo_{12}} are more stable than those of \ce{YFe_{12}}, and the $\Delta E$ value is decreased by 11.8\,meV/atom for type-I and by 5.7\,meV/atom for type-II. 
Although the three structures  of \ce{YCo_{12}} have much higher $T_{\text{C}}$ values than those of \ce{YFe_{12}}, 
the $M$ value is slightly decreased and the $T_{\text{C}}$ value is almost unchanged by the structural transformations in \ce{YCo_{12}}. 
These results suggest that the enhancement of $M$ and $T_{\text{C}}$ by the transformation into the monoclinic structures is a peculiar characteristic in the Fe-based system. 

\begin{figure}
\includegraphics[width=8.0cm]{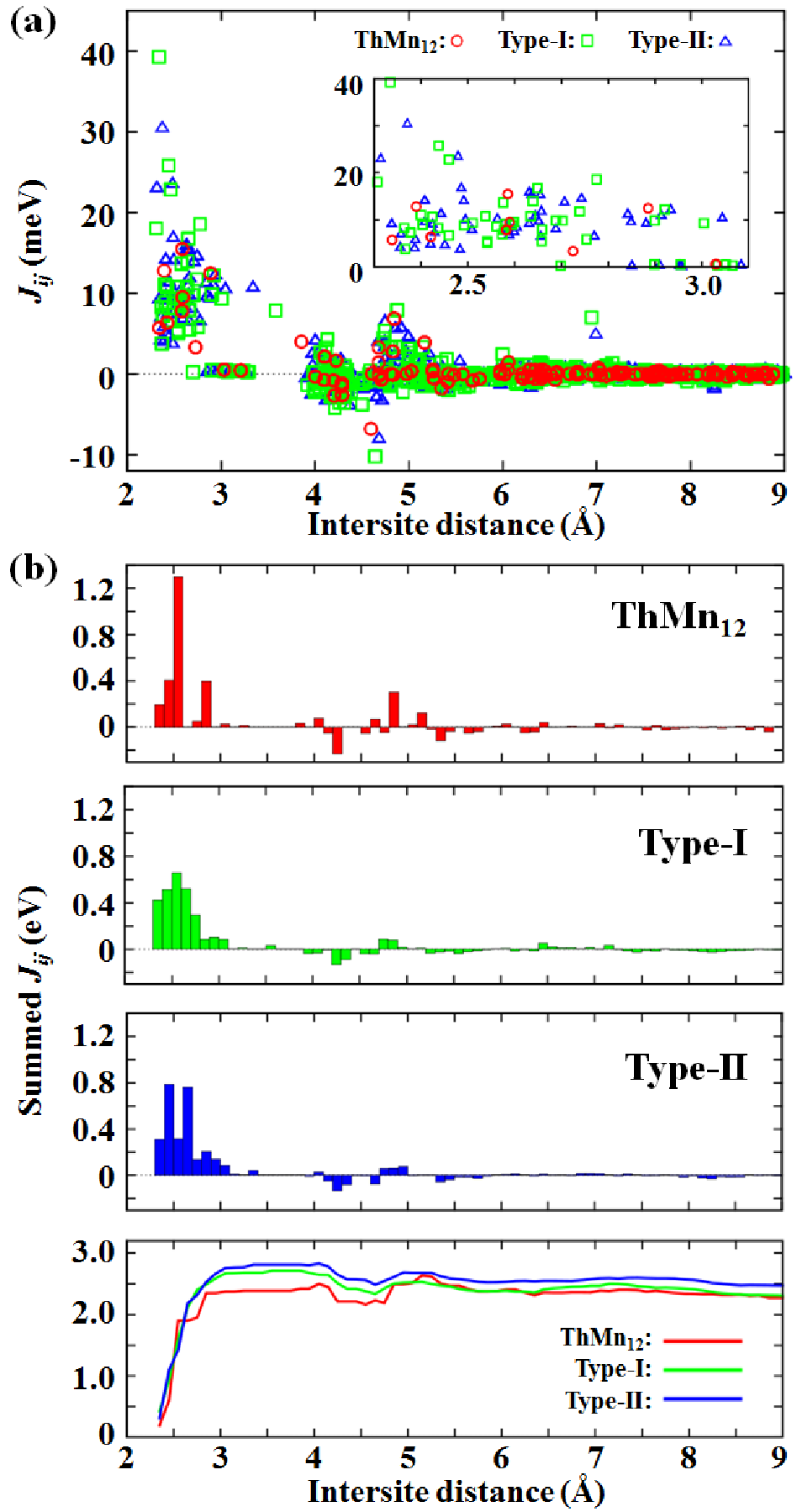}
\caption{\label{Fig-Jij} 
(Color online) (a) Magnetic coupling constant $J_{ij}$ for the intersite distance with respect to the \ce{ThMn_{12}}, type-I, and type-II structures in \ce{YFe_{12}}. (b) 
Summation of $J_{ij}$ values in the bins of the distance with the interval of 0.1\AA\, and the values integrated from zero distance.
}
\end{figure}
To see how the enhancement of $T_{\text{C}}$ in \ce{YFe_{12}} is caused in terms of their magnetic couplings, 
we calculated the magnetic coupling constant, $J_{ij}$, between the $i$th and $j$th sites for the \ce{ThMn_{12}}, type-I, and type-II structures. 
Figure \ref{Fig-Jij} shows (a) $J_{ij}$ as a function of the intersite distance in the range of 2--9\AA\, and (b) the summation of $J_{ij}$ values in the bins of the distance with the interval of 0.1\AA\, and the values integrated from zero distance. 
In the range of 2.3--2.6\AA, as shown in the inset of Fig.~\ref{Fig-Jij} (a), the type-I and type-II structures have $J_{ij}$ values larger than 20\,meV, which are absent in \ce{ThMn_{12}}.
However, as shown in the bottom panel in Fig.~\ref{Fig-Jij} (b), the $J_{ij}$ values integrated up to 2.6{\AA} of type-I and type-II are comparable with that of  \ce{ThMn_{12}}. 
Note that 
due to the high symmetry of the \ce{ThMn_{12}} structure, there are a number of symmetrically equivalent bonds that are shown duplicately
in Fig.~\ref{Fig-Jij} (a), which explains the hight of the 2.5--2.6{\AA} bin in the top panel of Fig.~\ref{Fig-Jij} (b).
In contrast, for type-I and type-II, modestly strong couplings appear broadly over 2.3--3.1\AA, which results in that the $J_{ij}$ values integrated up to 3.1{\AA} of type-I and type-II are larger than that of \ce{ThMn_{12}}. 
In the range longer than 3.1\AA, there emerge positive and negative $J_{ij}$ values, which contribute to the increase and decrease of $T_{\text{C}}$, respectively. 
Correspondingly, the integrated values have fluctuation as functions of the distance, which is clearly visible in the range of 4--5\AA. The fluctuation is suppressed and the integrated values exhibit converging behavior in the region of longer distances.

\begin{figure}
\includegraphics[width=8.6cm]{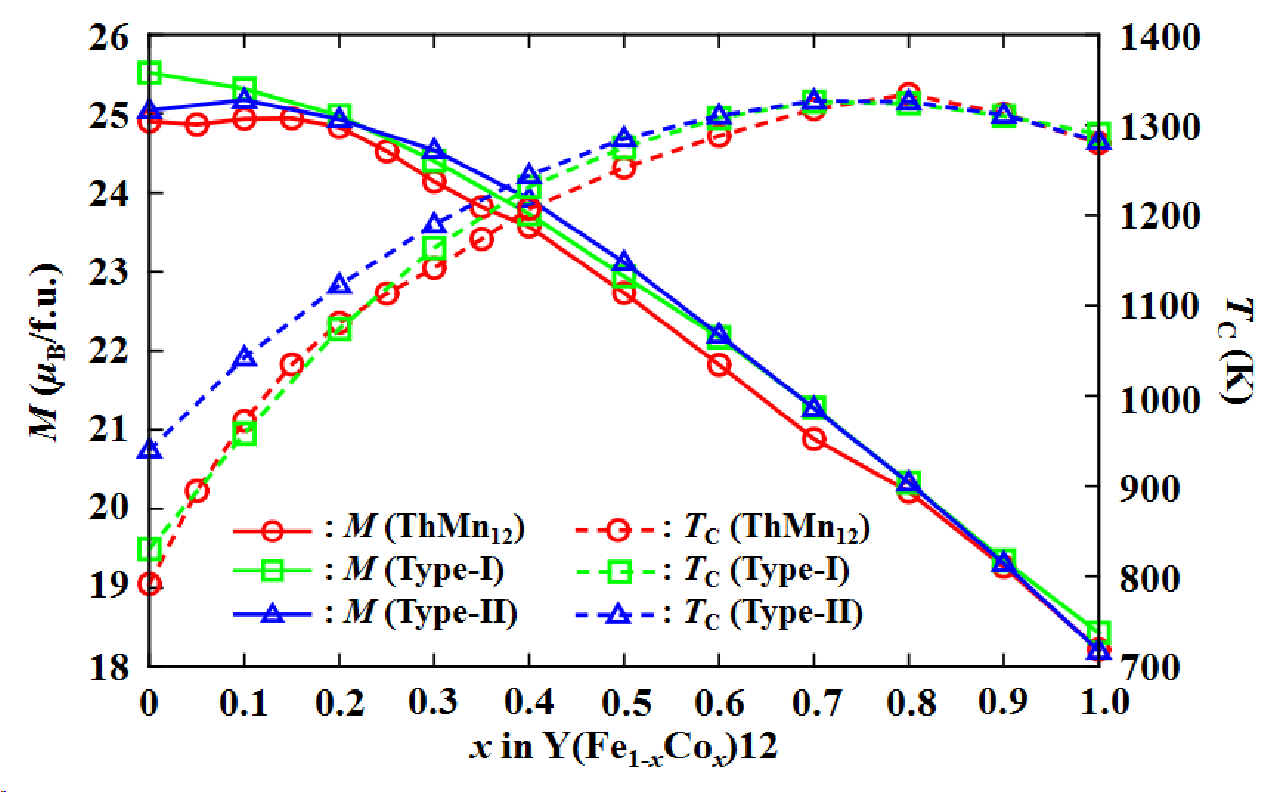}
\caption{\label{Fig-YFeCo} 
(Color online) Variations of magnetization $M$ and Curie temperature $T_{\text{C}}$ for Co concentration $x$ in the pseudo-binary compound \ce{Y(Fe_{1-$x$}Co_{$x$})_{12}}.}
\end{figure}
We next investigated the enhancement of $M$ and $T_{\text{C}}$ in the pseudo-binary system, \ce{Y(Fe_{1-$x$}Co_{$x$})_{12}}. 
Co has commonly been used for the enhancement of magnetic properties at finite temperatures, especially $T_{\text{C}}$, in Fe-based systems.@
In addition, the Co inclusion is worth investigating in terms of the enhancement of the $M$ value at zero temperature.\cite{Hasegawa1971,Hasegawa1972,Kanamori1990,Dederichs1991}
Figure \ref{Fig-YFeCo} shows values  of $M$ and $T_{\text{C}}$ as functions of $x$ in \ce{Y(Fe_{1-$x$}Co_{$x$})_{12}} for \ce{ThMn_{12}}, type-I, and type-II, which were calculated using AkaiKKR. 
The lattice parameters were fixed to the values at $x = 0$ for all the calculations. 
We confirmed that, although the $M$ values at $x = 0$ and 1 are different from those by QE shown in Table \ref{data_comparison}, the results are qualitatively consistent with each other. 
In the $x$ range of 0--0.7, as in the case of binary \ce{YFe_{12}}, the $M$ and $T_{\text{C}}$ values are increased by the structural transformations. 
For \ce{ThMn_{12}} and type-II, the Co-doping up to $x = 0.2$ increases the $T_{\text{C}}$ values while keeping the $M$ values at $x = 0$.
Notably, some Co-doped type-II systems have advantage both in $M$ and $T_{\text{C}}$ over \ce{ThMn_{12}} doped with the same amount of Co: 
$M = 25.2$\,$\mu_{\text{B}}$/f.u. and $T_{\text{C}} = 1042$\,K for \ce{Y(Fe_{0.9}Co_{$0.1$})_{12}} and 25.0\,$\mu_{\text{B}}$/f.u. and 1123\,K for \ce{Y(Fe_{0.8}Co_{$0.2$})_{12}} in type-II. 

\section{Conclusion}

We searched for the stable structures of \ce{YFe_{12}} using a crystal structure prediction technique based on the first-principles calculations and the genetic algorithm, and obtained two monoclinic $C2/m$ structures: type-I with the instability energy $\Delta E$ of 38.9\,meV/atom and type-II with 43.0\,meV/atom, in addition to the most stable \ce{ThMn_{12}} structure. These two $C2/m$ structures are related to the \ce{ThMn_{12}} structure by the permutation between the Y and Fe atoms. The total magnetization $M$ is increased from 25.6\,$\mu_{\text{B}}$/f.u. to 26.8\,$\mu_{\text{B}}$/f.u. for type-I and to 26.1\,$\mu_{\text{B}}$/f.u. for type-II. 
We calculated the Curie temperature $T_{\text{C}}$ for the two $C2/m$ structures within the mean-field approximation and obtained $T_{\text{C}}$ of 830\,K for type-I and 940\,K for type-II, which are both higher than that of \ce{ThMn_{12}}, 792\,K. 
In contrast, \ce{YCo_{12}} shows no enhancement of $M$ and $T_{\text{C}}$ caused by the structural transformations. 
We also calculated $M$ and $T_{\text{C}}$ for the pseudo-binary compound \ce{Y(Fe_{1-$x$}Co_{$x$})_{12}} and obtained the similar enhancements in the $x$ range of 0--0.7. 
Notably, $T_{\text{C}}$ is further increased without decrease of $M$ in the $x$ range of 0-0.2. 

\begin{acknowledgments}
This work was supported by the Ministry of Education, Culture, Sports, Science and Technology (MEXT) as ``The Elements Strategy Initiative Center for Magnetic Materials (ESICMM)'' (JPMXP0112101004) and ``Program for Promoting Researches on the Supercomputer Fugaku'' (DPMSD).
The computation was partly conducted using the facilities of the Supercomputer Center, the Institute for Solid State Physics, the University of Tokyo, the supercomputer of ACCMS, Kyoto University, and the HPCI System Research project (Project ID:hp200125).
\end{acknowledgments}

\end{document}